\documentclass[twocolumn,runningheads,natbib]{svjour2}
\smartqed  

\usepackage{graphicx}

\journalname{Astrophysics and Space Science}

\begin{document}

\title{Short time scale pulse stability of the Crab pulsar in the optical band\thanks{
This work has been supported by the Russian Foundation for Basic
Research (grant No 04-02-17555), Russian Academy of Sciences (program
"Evolution of Stars and Galaxies"), by the Russian Science Support
Foundation, and by INTAS (grant No 04-78-7366).
}}

\titlerunning{Pulse stability of the Crab pulsar}        

\author{S. Karpov \and
  G. Beskin \and
  A. Biryukov \and
  V. Debur \and
  V. Plokhotnichenko \and
  M. Redfern \and
  A. Shearer}

\authorrunning{S.Karpov et al} 

\institute{S. Karpov, G. Beskin, V. Debur \& V. Plokhotnichenko \at
              	Special Astrophysical Observatory of RAS \\
              	Nizhniy Arkhyz, Karachaevo-Cherkessia, Russia, 369167 \\
              	\email{karpov@sao.ru}       
	\and
        	A. Biryukov \at
	        Sternberg Astronomical Institute of MSU, \\ 
                13 Universitetsky pr., Moscow, 119992, Russia
        \and
	        M. Redfern \& A. Shearer \at
		National University of Ireland, Galway, \\
                University Road, Galway, Ireland
}

\date{Received: date / Accepted: date}

\maketitle

\begin{abstract}
The fine structure and the variations of the optical pulse shape and
phase of the Crab pulsar are studied on various
time scales. The observations have been carried out on
4-m William Hershel and 6-m BTA telescopes with APD photon counter,
photomultiplier based 4-channel photometer and PSD based panoramic
spectrophotopolarimeter with 1$\mu$s time resolution 
in 1994, 1999, 2003 and 2005-2006 years. 
The upper limit on the pulsar precession on Dec 2, 1999 is placed 
in the 10 s - 2 hours time range. 
The evidence of a varying from set to set fine structure of the main pulse
is found in the 1999 and 2003 years data. No such fine structure is detected
in the integral pulse shape of 1994, 1999 and 2003 years.

The drastic change of the pulse shape in the 2005-2006 years set is detected
along with the pulse shape variability and quasi-periodic phase shifts.

\end{abstract}

\section{Introduction}

Over the last 30 years the Crab pulsar has been extensively studied.
The reasons for it are clear -- it is the brightest pulsar seen in 
optics, it is nearby and young.
However, the post popular groups contemporary theories of the Crab high-energy 
emission, the ``polar cap'' \citep{daugherty} and ``outer gap'' \citep{cheng}
ones, can't explain the whole set of observational data.

One of the main properties of the Crab emission is the very high stability
of its optical pulse shape despite the secular decrease of the luminosity,
related to the spin rate decrease \citep{pacini,nasuti}.

At the same time
the pulsars in general and the Crab itself are unstable. The instabilities
manifest itself as a glitches, likely related to the changes of the
neutron star crust, timing noise, powered by the collective processes in the
superfluid internal parts of it, magnetospheric instabilities, results of
the wisps around the pulsar, precession, et al. All these factors may
influence the optical pulse structure and change it on various time scales,
both in periodic and stochastic way.

However,
it has been found early that the variations of the Crab optical light curve,
in contrast with the radio ones,
are governed by the Poissonian statistics \citep{Kristian}. 
A number of observations show the absence of non-stationary effects in
the structure, intensity and the duration of the Crab optical pulses, and the 
restrictions on the regular and stochastic fine structure of its pulse on the
time scales from 3$\mu$s to 500$\mu$s \citep{beskin,percival}, the fluctuations
of the pulse intensity \citep{Kristian}.

Along with the increase of the observational time spans and the accuracy of measurements
the small changes of the optical pulse intensity, synchronous with the giant radio
pulses, have been detected \citep{shearer}. Also, the evidence for the short time scale
precession of the pulsar has been detected by studying its optical light curve \citep{cadez}.

All this raises the importance of the monitoring of the Crab optical emission with
high time resolution.

\section{Observations}

\begin{table*}[t]
\caption{Log of observations}
\centering
\label{table_observations}
\begin{tabular}{lllll}
\hline\noalign{\smallskip}
Date & Telescope & Instrument & Duration, sec & Spectral range  \\[3pt]
\tableheadseprule\noalign{\smallskip}
Dec 7, 1994 & BTA, Russia & Four-color photometer & 2400 & U+B+V+R \\ 
 & & with photomultiplier & & \\
Dec 2, 1999 & WHT, Canary Islands & Avalanche photo-diode & 6600 & R \\
Nov 15, 2003 & BTA, Russia & Avalanche photo-diode & 1800 & R \\
Dec 29, 2005 -  & BTA, Russia & Panoramic spectro-polarimeter & 48000 & 4000 - 7000 A \\
Jan 3, 2006 & & with position-sensitive detector & & \\
\noalign{\smallskip}\hline
\end{tabular}
\end{table*}

We analyzed the sample of observational data obtained by our group over the
time span of 12 years on different telescopes. The details of
observations are summarized in Table \ref{table_observations}. The equipment used were
four-color standard photometer with diaphragms based on
photomultipliers, fast photometer with avalanche photo-diodes \citep{shearer}
and panoramic spectro-polarimeter based on position-sensitive detector \citep{psd,mppp}. 
All devices provide the 1$\mu$s time resolution.

For each data set the list of photon
arrival times has been formed. They have been processed in the same way by using the same 
software to exclude the systematic differences due to data analysis
inconsistencies. Photon arrival times have been corrected to the barycenter
of the Solar System using the adapted version of {\it axBary} code by 
Arnold Rots. The accuracy of this code has been tested with detailed examples
provided by \cite{lyne_bary_examples} and is found to be better than 2$\mu$s.

The barycentered photon lists then have been folded using both Jodrell-Bank 
radio ephemerides \cite{pjb_ephem} and our own fast-folding based method of timing model fitting.

The accuracy of timing model is proved to be better than at least several 
microseconds (see Figure \ref{fig_shift_1999}),
which permits to fold the light curve with 5000 bin (6.6 $\mu$s)
resolution.

\section{Phase stability}
\begin{figure}
\centering
{\centering \resizebox*{1\columnwidth}{!}{\includegraphics{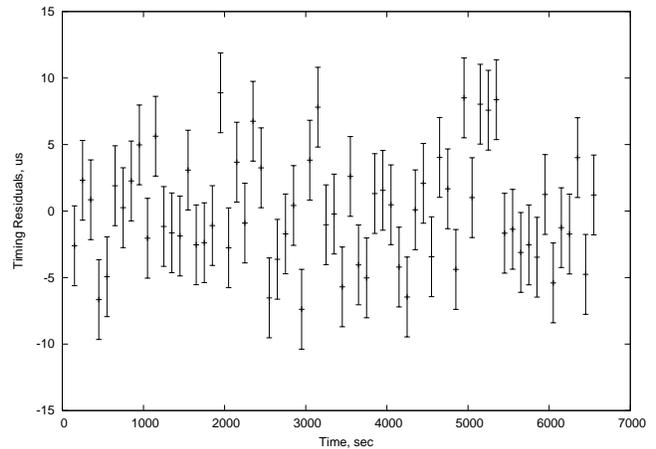}} \par}
\caption{Timing residuals of the Crab pulsar after applying second-order timing model
(up to second frequency derivative). It corresponds to the Gaussian noise with 4.1 $\mu$s rms.}
\label{fig_shift_1999}
\end{figure}

\begin{figure}
\centering
{\centering \resizebox*{1\columnwidth}{!}{\includegraphics[angle=270]{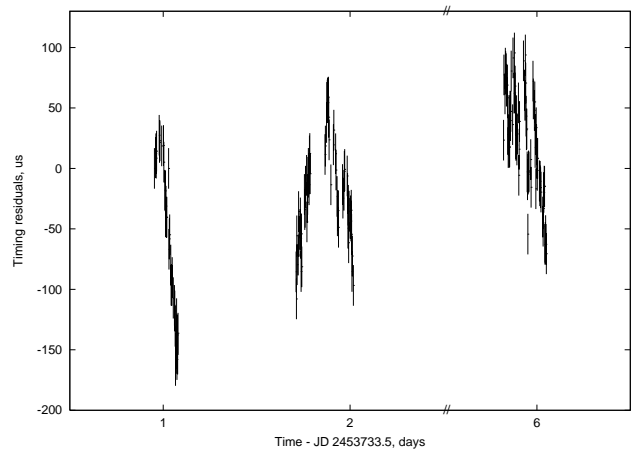}} \par}
\caption{Timing residuals of the Crab pulsar after applying second-order timing model. 
The quasi-periodic behaviour with characteristic time scale of 0.7 days is seen.}
\label{fig_shift_2006}
\end{figure}

We performed the search for timing model residuals 
using two longest continuous data sets of 1999 and 2005-2006 years. The data
has been divided into the number of subsets of fixed length and they have been
folded separately using the same base epoch. Then the sample light curves have
been cross-correlated with the standard one (which has been derived for each set
separately by folding the whole data) and its phase shift have been derived by
fitting the maximum of the cross-correlation function with the Gaussian.
The results for 1999 year set are shown in Figure \ref{fig_shift_1999}. No
evidence for significant deviations from zero is seen, the phase is consistent
with the Gaussian noise with 4.1$\mu$s rms in the 10 s - 2 hr time range.

The data of the last set of 2005-2006 years, however, show the significant
quasi-periodic variations with $\sim 2.5\cdot10^{-3} P$ rms amplitude.
The characteristic time scale of the variations is estimated to be 
roughly $0.7$ d.

\section{Pulse shape}

\begin{figure}
\centering
{\centering \resizebox*{1\columnwidth}{!}{\includegraphics{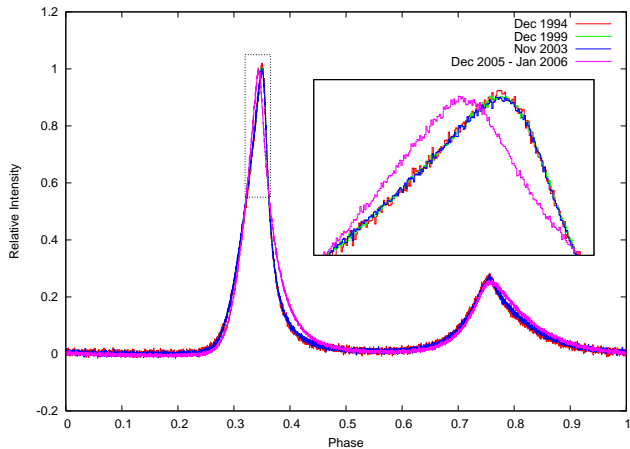}} \par}
\caption{Phased light curves of the Crab pulsar for all observations sets, scaled to the same pulse height.}
\label{fig_shape_all}
\end{figure}

\begin{figure}
\centering
{\centering \resizebox*{1\columnwidth}{!}{\includegraphics[angle=270]{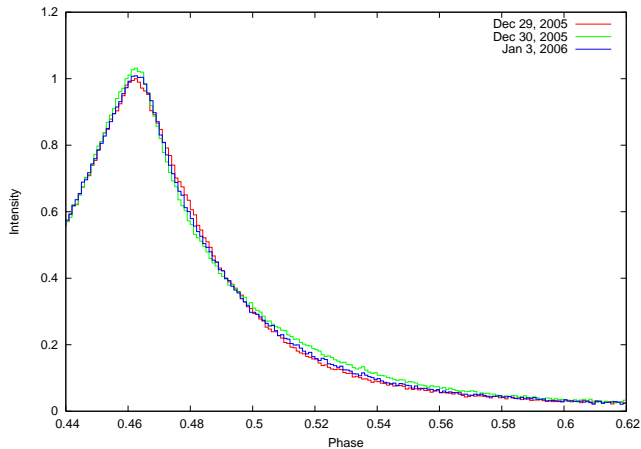}} \par}
\caption{Phased light curves of the Crab pulsar for the three nights of the Dec 2005 - Jan 2006 set.}
\label{fig_shape_nights}
\end{figure}

Due to the presence of significant residuals relative to the timing model
the pulse profile during the observations of 2005-2006 years can't be derived
by folding the whole data set directly. Instead, we divided the data set into 
the one-hour segments and folded them separately applying the time shift corrections
to compensate the phase residuals. The intrinsic phase shift inside each block is
less than $2\cdot10^{-4}$, so the folding with 5000 bins is possible.
The folded light curves have been co-added.

All the other data has been folded directly and shifted in phase to the same pulse position
for the ease of comparison. All pulse profiles are shown in
Figure \ref{fig_shape_all}, with the off-pulse emission subtracted and pulse height scaled to 
the same value.

The profiles of 1994, 1999 and 2003 years are in a perfect agreement with each other.
The profile of 2005-2006 years, however, deviates from them significantly -- the pulse
remains of the same FWHM while its skewness is much smaller, and its shape is
nearly symmetric. 

We folded the data of this set for each of three observational nights separately using the
same method. These profiles are shown in Figure \ref{fig_shape_nights}. There is the
significant variation of its shape from night to night. Unfortunately the low
amount of data available do not permit to track the profile shape change inside each
night and check whether it is smooth or whether the shape is correlated with the
timing residuals.

\section{Pulse fine structure}
\begin{figure}
\centering
{\centering \resizebox*{1\columnwidth}{!}{\includegraphics{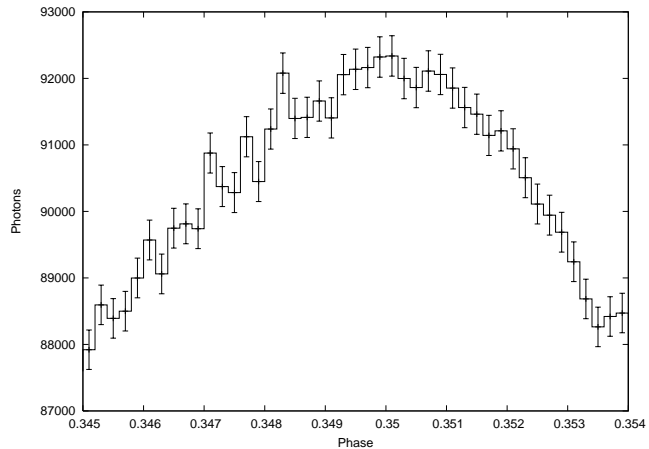}} \par}
\caption{The main pulse peak of the sum of light curves of 1994, 1999 and 2003 years data.}
\label{fig_pulse_123}
\end{figure}

\begin{figure}
\centering
{\centering \resizebox*{1\columnwidth}{!}{\includegraphics{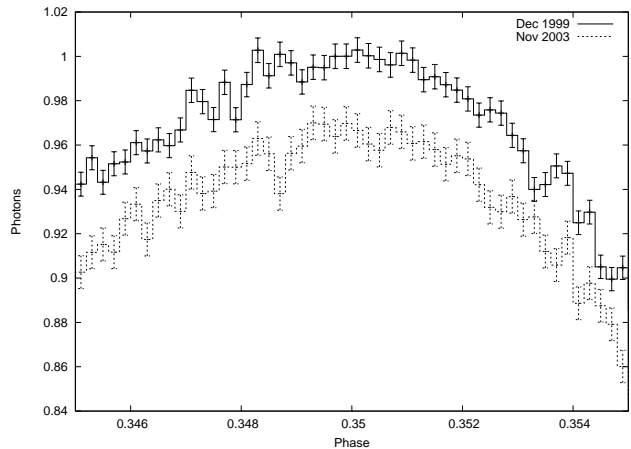}} \par}
\caption{The comparison of the peaks of 1999 and 2003 years. The peaks are shifted vertically  for 0.03 for clearance.}
\label{fig_pulse_23}
\end{figure}

For the first three observational set where the pulse profile is stable we 
performed the search for the fine structure of the main pulse. The data sets
has been reduced to the same phase base point with precision better than half
of the phase bin (less than 3.3$\mu$s) and the cumulative light curve
has been computed. The peak region of it is shown on Figure \ref{fig_pulse_123}.
No statistically significant deviations from the smooth peak
shape is seen. However, light curves of 1999 and 2003 years data sets
alone (plotted on Figure \ref{fig_pulse_23}) each show the evidence of 
fine structure on the level of 3-5 sigma
(roughly 1 \% of the intensity) with typical duration of 10-30
$\mu$s. Such details may give an evidence of coherent generation of
optical emission, if the emission generation region is deep enough
(deeper than 0.1 of light cylinder radius), due to brightness
temperature exceedes $10^{12}$ K.

\section{Discussion}

The results of the last data set differs significantly from all
previous results.  We studied carefully the possibility of its being
the result of some hardware or data processing problem.

The data of last set has been acquired using the panoramic
spectro-polarimeter based on the position-sensitive photon counter \citep{psd,mppp} in
low-resolution spectral mode. There is no difference of the pulse profile 
in different spectral bands (derived using different parts of the spectrum
detected).
The detector behaviour is proved to be linear in the flux range used.
The data acquisition system, ``Quantochron 4-48'', which records the
time of arrival of the detected photons, has been checked for short time scale
stability by recording the signal from the stationary 100-Hz generator.
It has been processed and folded in the same way as a pulsar one (passing the unnecessary
barycentering step). It shows no distortion of the signal shape
larger than 1$\mu$s. Large scale timing stability of acquisition system is
ensured by means of 1 Hz and 10 kHz frequency signals from GPS
receiver. The barycentering correction code passed the tests provided
in \cite{lyne_bary_examples} with the accuracy of 1 $\mu$s.
The correctness of the radio ephemerides has been checked by performing the timing
model fitting using our own software, the results of phase shift
and folding analysis agree with ones based on the radio data.
There is no small time scale (of order of 100 seconds and larger)
changes of the pulse profile inside the set with amplitude comparable to
the difference between the last set and previous ones.

Taking into account all these arguments we may conclude that the pulse profile
change and quasi-periodic phase shifts detected in this observational set
is most likely not related to the hardware or software problems of the
equipment used.

The detection of the variations of both the pulse arrival times and its shape
strongly supports the geometrical interpretation of the effect. It may be
described as a quasi-periodic change of the pulsar beam orientation due
to the strong precession commenced suddenly before the observations, but after
the previous set. It may be related to the very strong glitch of the Feb-Mar 2004,
or other recent change in the neutron star state \citep{lyne_bary_examples}

\section{Conclusions}

We analyzed the data of several sets of optical observations with high
temporal resolution of the Crab pulsar performed by our group 
over the 12 last years. 

No evidence for short time scale precession (like 60-sec free precession
discovered in \cite{cadez} ) is detected on the
level of $10^{-5}$ - $10^{-7}$ s$^{-1}$ pulsar frequency variation on 10 s - 2 hours
time scale on Dec 2, 1999 (see Figure \ref{fig_shift_1999}), which corresponds 
to the precession wobble angle to be
less than approximately $2\cdot10^{-3}$. Also, no signatures of short time scale
timing noise is seen in this data set.

No significant fine structure is detected in the integral pulse profile
of 1994, 1999 and 2003 years data set (see Figure \ref{fig_pulse_123}), however, 
each data set alone show the evidence of fine structure on the level of 3-5 sigma,
which may be related to its instability on the time scale of years along with 
the stability of the pulse shape on the same scale.

We discovered the significant change of the time-averaged Crab pulse profile in the
Dec 2005 - Jan 2006 set of observations. The pulse profile also shows the variations 
between the nights. Also, the quasi-periodic phase shifts in respect to the
second-order timing solution (up to second frequency derivative) has been
detected in the data with amplitude of $\sim 100\mu$s and characteristic time scale
of 0.7 days. We have not found any hardware or software issue able to mimic
such pulsar behaviour. These results may be interpreted as a geometric effects
due to the Crab precession suddenly started between our observations of 2003 and 2005-2006 years.


\begin{thebibliography}{99}

\bibitem[\protect\citeauthoryear{Beskin et al.}{1983}]{beskin}
Beskin G. M., Neizvestnyi S. I., Pimonov A. A. et al. 
Sov.Astron.Lett {\bf 9}, 148 (1983)

\bibitem[\protect\citeauthoryear{Cadez et al.}{2001}]{cadez}
Cadez A., Vidrih S., Galieie M. et al. A\&A, {\bf 366}, 930 (2001)

\bibitem[\protect\citeauthoryear{Cheng et al.}{2000}]{cheng}
Cheng K.S., Ruderman M. \& Zhang L. ApJ {\bf 537}, 964 (2000)

\bibitem[\protect\citeauthoryear{Daugherty \& Harding}{1996}]{daugherty}
Daugherty J.K. \& Harding A.K. ApJ {\bf 458}, 278 (1996)

\bibitem[\protect\citeauthoryear{Debur et al.}{2003}]{psd}
Debur V., Arkhipova T., Beskin G. et al.
Nuclear Instruments and Methods in Physics Research, {\bf A 513}, 127 (2003)

\bibitem[\protect\citeauthoryear{Jordan}{2006}]{jb_ephem}
Jordan C.A.: Private communication. (2006)

\bibitem[\protect\citeauthoryear{Kristian et al.}{1970}]{Kristian}
Kristian J., Visvanathan N., Vestphal J.A. et al. ApJ {\bf 162}, 475 (1970)

\bibitem[\protect\citeauthoryear{Lyne et al.}{2005}]{lyne_bary_examples}
Lyne A.G., Jordan C.A. \& Roberts M.E. Crab Monthly Ephemeris available on
http://www.jb.man.ac.uk/~pulsar/crab.html (2005)

\bibitem[\protect\citeauthoryear{Nasuti et al.}{1996}]{nasuti}
Nasuti F.P., Mignani R., Caraveo P.A. et al. A\&A {\bf 314}, 849 (1996)

\bibitem[\protect\citeauthoryear{Pacini}{1971}]{pacini}
Pacini F. ApJ {\bf 163}, 17 (1971)

\bibitem[\protect\citeauthoryear{Percival et al.}{1993}]{percival}
Percival J.W., Biggs J.D., Dolan J.F. et al. ApJ {\bf 407}, 276 (1993)

\bibitem[\protect\citeauthoryear{Plokhotnichenko et al.}{2003}]{mppp}
Plokhotnichenko V., Beskin G., Debur V. et al. 
Nuclear Instruments and Methods in Physics Research, {\bf A 513}, 167 (2003)

\bibitem[\protect\citeauthoryear{Shearer et al.}{2003}]{shearer}
Shearer A., Stappers B., O'Connor, P. et al. Science, {\bf 301}, 493 (2003)

\end{thebibliography}
\end{document}